\documentclass[preprint2]{aastex}

\newcommand{\twco}{$^{12}$CO}
\newcommand{\thco}{$^{13}$CO}

\shorttitle{THE NEARBY QSO I~ZW~1}
\shortauthors{Staguhn et al.}

\begin{document}

\title{Resolving the Host Galaxy of the Nearby QSO I~Zw~1 with
Sub-Arcsecond Multi-Transition Molecular Line Observations\altaffilmark{1,2}}


\author{J.G. Staguhn\altaffilmark{3,4,5},  E. Schinnerer\altaffilmark{6,7}, A. Eckart\altaffilmark{8}, J. Scharw\"achter\altaffilmark{8}}

\email{staguhn@stars.gsfc.nasa.gov}




\altaffiltext{1}{Based on observations carried out with the Berkeley
Illinois Maryland Association (BIMA) observatory. The BIMA
observatory is supported by NSF grant AST-9981289}
\altaffiltext{2}{Based on observations carried out with the IRAM
Plateau de Bure Interferometer. IRAM is supported by INSU/CNRS
(France), MPG (Germany) and IGN (Spain)}
\altaffiltext{3}{NASA/Goddard Space Flight Center, Greenbelt, MD 20771, USA}
\altaffiltext{4}{Department of Astronomy, University of Maryland, College Park, MD 20742, USA}
\altaffiltext{5}{SSAI, Lanham, MD 20706, USA}
\altaffiltext{6}{NRAO, Socorro, NM 87801, USA}
\altaffiltext{7}{Jansky Postdoctoral Fellow at the National Radio Astronomy Observatory}
\altaffiltext{8}{1. Physikal. Inst., Univ. zu K\"oln, 50923 K\"oln, Germany}

\begin{abstract}

We present the first sub-kpc $(\sim0.7''\approx 850$ pc$)$ resolution
\twco(1-0) molecular line observations of the ISM in the host galaxy
of the QSO I~Zw~1. The observations were obtained with the BIMA
mm-interferometer in its compact A configuration. The BIMA data are
complemented by new observations of the \twco(2-1) and \thco(1-0) line
with IRAM Plateau de Bure mm-interferometer (PdBI) at $0.9''$ and $1.9''$
resolution, respectively. These measurements, which are part of a
multi-wavelength study of the host galaxy of I~Zw~1, are aimed at
comparing the ISM properties of a QSO host with those of nearby galaxies
as well as to obtain constraints on galaxy formation/evolution models.
Our images of the \twco(1-0) line emission show a ring-like structure
in the circumnuclear molecular gas distribution with an inner radius of about
1.2~kpc. The presence of such a molecular gas ring was predicted from
earlier lower angular resolution PdBI \twco(1-0) observations. A
comparison of the BIMA data with IRAM PdBI \twco(2-1) 
observations shows variations in the excitation conditions of the molecular gas in
the innermost $1.5''$ comprising the nuclear region of I~Zw~1. The
observed properties of the molecular cloud complexes in the disk of
the host galaxy suggest that they can be the sites of massive
circumnuclear star formation, and show no indications of excitation by
the nuclear AGN. This all indicates that the molecular gas in a QSO
host galaxy has similar properties to the gas observed in nearby low 
luminosity AGNs.

\end{abstract}




\section{Introduction}

One of the open questions in galaxy evolution is concerned with the
formation of active galactic nuclei (AGN) and its relation to star
formation (SF) in early type galaxies. The separation of starburst
and AGN components in extragalactic objects
-- especially in host galaxies of quasars and QSOs -- is a key problem
in the investigation of evolutionary sequences proposed for AGNs
\citep{nor88,san88,rie88,haa03}. Although it is not exactly known how 
the host galaxy affects the energy release of the QSO, there is
statistical evidence for preceding mergers or current interaction of
QSO host galaxies with a companion galaxy \citep{mcl94,lim99}.
It is believed that quasar activity is a common, but
short lived, phenomenon in galaxy evolution \citep{mcl99}.
Furthermore, host galaxies of QSOs show enhanced SF activities
\citep{cou98}. The transition between Ultra Luminous Infrared Galaxies 
(ULIRG) and AGN seems to be
continuous, all showing signs of enhanced star formation in their
nuclear regions \citep{gen98,mcl99}.

Millimeter molecular line observations are ideal for studying the
mechanisms which transport the molecular gas into the
AGN. Observations of abundances, excitation and
dynamics of the molecular interstellar medium in the central regions
of AGN are essential, since the interstellar matter provides the
''fuel'' for star formation as well as the central engine.
Consequently, there are large numbers of high resolution
observations available for nearby objects. These observations have
revealed the presence of circumnuclear starburst rings in a large
number of (nearby) active and IR luminous galaxies \citep [e.g. NGC 1068,] [] {pla91, gen95, hel03}. 
The extension of high angular resolution
observations of molecular gas emission lines to QSO hosts is
imperative in order to understand the connection between local active galaxies
and high-z QSOs. The results can be used to refine model predictions
of the physical conditions in high z- QSOs \citep[such as used e.g. in] [] {com99}. 

Due to the limited angular resolution of single dish mm-wavelength
telescopes,
only interferometric observations at these wavelengths allow insight
into the morphology and into the kinematics of molecular clouds in the
nuclear region of QSO host galaxies with sufficient angular
resolution. High resolution molecular line observations allow for detailed
kinematic studies of the cold interstellar medium and the
derivation of important parameters such as gas masses, surface mass
densities and star formation efficiencies. Observations of multiple CO transitions
reveal excitation conditions and thus can be used to
constrain the physical conditions in an observed region.
They can be used to better constrain
the  contribution of star formation to the total observed infrared
luminosities.


\subsection{The nearby QSO I~Zw~1} \label{IZw1_Intro}

The radio-quiet QSO I~Zw~1 is regarded as the closest QSO which can be
used for detailed studies of its host galaxy
(Tab. \ref{tab:izw1}). I~Zw~1 has a  systemic velocity of 18,290 km s$^{-1}$, which corresponds
to a redshift of 0.0611 \citep{con85}, or a distance of 255 Mpc \citep[$h_0=0.72$,] [] {spe03}. 
The nucleus of I~Zw~1 is extremely bright in the optical \citep[$M_B$
of -23.45 mag,] [] {sch83, bar89}) and also has very bright X-ray emission \citep{kru90}. 
 I~Zw~1 belongs to the class of infrared luminous galaxies
\citep[$L_{FIR}=10^{11.3}L\sun$,] [] {haa03}, however its QSO nature
is not apparent in the FIR.  There is a  strong indication for interaction between 
I~Zw~1 and a neighboring companion \citep[e.g.] [] {lim99,schar03}. 

A direct comparison between the FIR emission and CO molecular line
emission in the host galaxy reveals that the FIR continuum emission of
I~Zw~1 is predominantly thermal in nature \citep{bar89}. The FIR to 
\twco(1-0) ratio shows a star formation
efficiency of about 30 $L_\sun/M_\sun$ in the nuclear region \citep{eck94}. \citet{schin98} 
 show that the molecular gas
mass in the central $3.3''$ (3.7 kpc) is 2/3 of the total observed mass
of $9 \times 10^9 M_\sun$. A kinematic analysis of their 1.9'' resolution PdBI \twco(1-0) observations suggests that the
central molecular gas is distributed in a ring of $1.5''$ ($\sim 1.9$
kpc) diameter.  However the observations do not spatially resolve this ring. High resolution NIR observations show that the nuclear
NIR spectrum is dominated by emission from the AGN, nevertheless about
25\% of the NIR emission can be attributed to a nuclear starburst
\citep{schin98},  which might be located in the molecular gas
ring.  The  star formation activity in the host galaxy of I~Zw~1
underlines the important role of starbursts in the evolution of QSOs.

Only sub-arcsecond molecular line observations would allow actual 
imaging of this possible starburst ring. The extension of BIMA's longest
baselines up to a maximum of almost 2 km makes the array the
millimeter instrument with the highest angular resolution currently
available. Here we present \twco(1-0) $\sim 0.7''$ angular resolution
observations of the QSO I~Zw~1. This corresponds to a spatial
resolution of roughly 850~pc. We also present new observations with
the IRAM PdBI in \twco(2-1) and \thco(1-0) line at $0".9$ and $2.''1$
resolution, respectively.

\begin{deluxetable}{lcc}
\tablecaption{Adopted Properties of I~Zw~1 (=PG0050+124) \label{tab:izw1}}
\tablewidth{0pt}
\tablehead{
\colhead{Property} & \colhead{Value}}
\startdata
Right ascension (J2000.0) & 00$^h$53$^m$34$^s$.9\\
Declination (J2000.0) & +12$\degr 41'36''.2$\\
Inclination (deg) & 38\\
Position angle (deg) & 135\\
Systemic velocity (km s$^{-1}$) & 18,290\\
Distance (Mpc)& 255\\
$1''$ & 1.24 kpc
\enddata
\tablenotetext{-}{The coordinates refer to the phase center of the millimeter observations. The
inclination and position angle are taken from \citet{schin98}, while the systemic
velocity is from \citet{con85}. }
\end{deluxetable}


\section{Observations and data reduction}

\begin{figure}
\plotone{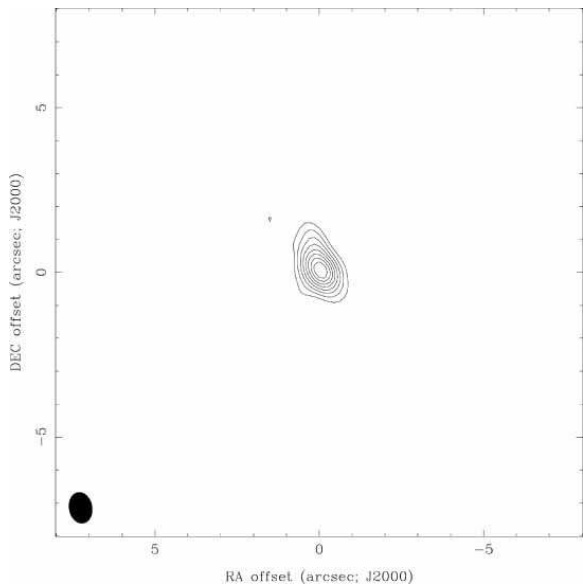}
\caption{BIMA image of the test source 0121+118 in \twco(1-0). The contour lines start at
20\% of the peak level 
in increments of 10\%. 
Only the 20\% contour shows a slight deviation from the ideal point
source response. The synthesized beam is shown in the lower left of the image.} 
\label{fig1}
\end{figure}
\begin{figure}
\plotone{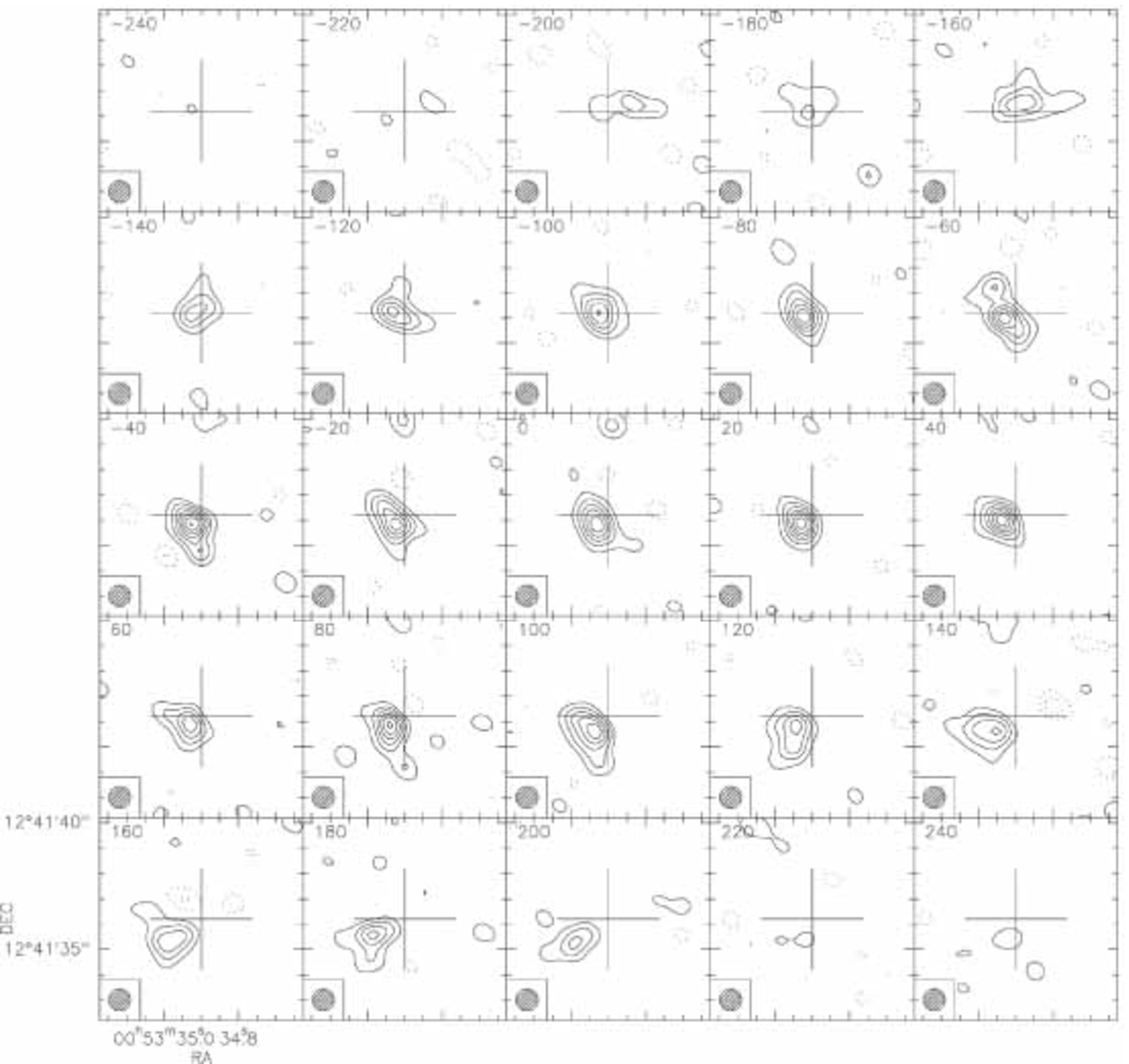}
\caption{Channel maps of \twco(2-1) line emission in I~Zw~1 observed with the  PdBI.
The relative velocities with respect to  the observed systemic velocity are
indicated in the upper left corner of each panel.
The data have a resolution of $0.86''$ and the contours are in steps of
5\,mJy beam$^{-1}$ ($\sim 3\sigma$). The rotation of the galactic CO disk is
clearly evident in the channel maps. The clean beam is shown for reference in the lower left corner of each panel.}
\label{fig:co21channel}
\end{figure}

\subsection{BIMA observations}

The data presented here were taken in December 1999 in the
compact A array configuration (A-) of the ten element BIMA array \citep{wel96}
with baselines between 80 and 1310~m. Four more tracks of A array observations were
obtained in February 2000, December 2001, and January 2002, but all of
those yielded only poor quality data, demonstrating that this project
poses requirements which are at the limit of what can be done with
long baseline mm-interferometry to-date. The digital correlator was
configured to cover two 800 MHz bands with the upper sideband centered
at the redshifted \twco(1-0) frequency of 108.633 GHz.
The sky opacity is significantly lower at the frequency of
redshifted \twco(1-0) transition than at the rest frequency.
System temperatures during the observations ranged from 180 to
400~K, single-sideband. We flagged data such that most observations
with system temperatures above 300~K were rejected. The data were reduced 
with the MIRIAD software package \citep{sau95}.

The fluctuations in the atmosphere follow a Kolmogorov power law
distribution up to the maximum baseline of ~ 2 km in the A array
configuration of the BIMA array. This means that phase variations
increase with baseline length, and therefore long baseline observations
require a particularly careful phase calibration. We employed fast
phase referencing: the observations were switched between the source,
the phase calibrator (0108+015, distance from I~Zw~1 $\sim 11^o$) and
an additional point source, the quasar 0121+118 (angular distance from
I~Zw~1 $\sim 7^o$) which serves as a test source to determine the
accuracy of the solutions for the phases. One complete cycle takes
less than two minutes in order to follow closely the atmospheric
phase \citep{hol93}.

The test calibrator was used to select the best quality data: only
observing cycles for which the test source maps to a point source down
to the 20\% level of its peak flux were
selected (Fig.~\ref{fig1}). This criterion was chosen to allow for
a good quality image of the source. The reverse check, using 0121+118
as phase calibrator and 0108+015 as the test source, verified that the
phase solutions derived from the test calibrator are of the same
accuracy. Only the data taken in December 1999 have 
sufficient phase coherence to yield sub-arcsecond resolution
observations. The data presented here are only from this observing
track. The phase solutions derived from the test source 0121+118 were used for
the images shown here, since this source is closer to I~Zw~1 and
therefore provides the best phase coherence. 0121+118 has a very flat
spectrum with a flux of 1.1 Jy at 3.7~cm, 2~cm, and 7~mm (VLA
calibrator list) and we use the same value for our 3~mm band. We
estimate the uncertainty in the amplitude calibration to be better
than 30\%. In order to increase the signal to noise ratio in the data,
we smoothed them to an effective angular resolution of $0.7''$.

\subsection{IRAM PdBI observations}

I~Zw~1 was observed simultaneously in the \twco(2-1) and \thco(1-0)
line in August 1996 with the four-element IRAM Plateau de Bure
millimeter interferometer (PdBI) in the C configuration. Further
dual-frequency observations in the B configuration were obtained in
January 1997 and January/February 1998 with the then five-element
array. The correlator was centered on the redshifted \twco(2-1)
frequency of 217.26321\,GHz at 1~mm and the redshifted \thco(1-0)
frequency of 103.85578\,GHz at 3~mm. With baselines between 40 and
280~m the data have a spatial resolution of $\sim 0.86''$ with
uniform weighting for the \twco(2-1) line (Fig. \ref{fig:co21channel}),
and $\sim 2.1''$ with natural weighting for the \thco(1-0) line. The
noise per 20(40) km s$^{-1}$ wide channel is $\rm 1.7(0.33)$ mJy beam$^{-1}$ in the
combined data for the \twco(2-1)(\thco(1-0)) line. The quasar 3C454.3
served as a passband calibrator, while standard IRAM PdBI phase calibrators were
observed every 20 minutes. The data were calibrated and mapped using
the IRAM GILDAS software package \citep{gui00}.

\section{Results}

\begin{figure}
\plotone{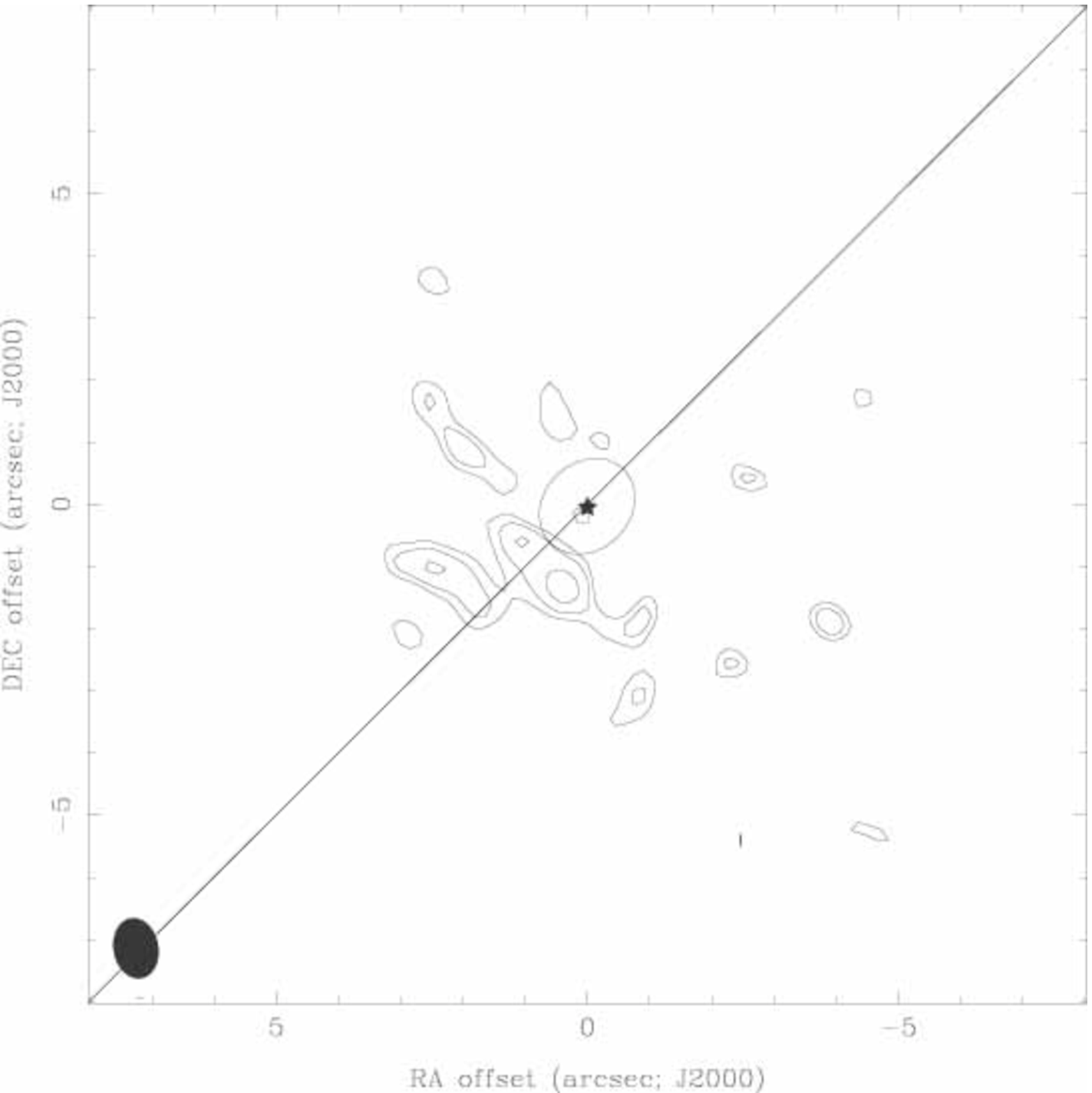}
\caption{BIMA image of the \twco(1-0) emission smoothed to a resolution of $0.7''$. The solid contours show the velocity integrated emission between $-160$ and  240 km s$^{-1}$ in $1\sigma$ steps (3\,mJy~beam$^{-1}$) starting at $4\,\sigma$ (9 mJy~beam$^{-1}$). If present, dashed contours would represent the corresponding negative values for the flux density. The ellipse in the center of the image marks the projected distance of 1~kpc ($0.8''$) from the center. The diagonal line indicates the major kinematic axis of the galaxy. The ellipse in the lower left corner shows the synthesized beam.} 
\label{fig2}
\end{figure}
\begin{figure}
\plotone{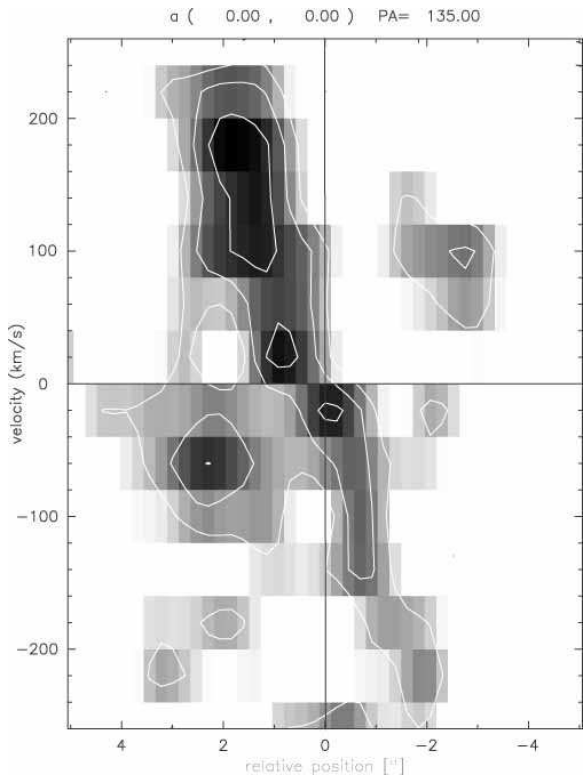}
\caption{ BIMA  \twco(1-0) position-velocity diagram along the major axis ($135\degr$) of I~Zw~1 with a spatial resolution of $0.7''$. The contours start at 60\% (3 $\sigma$) of the peak flux of 35 mJy~beam$^{-1}$ with increments of 15\% (1$\sigma$).} 
\label{fig3}
\end{figure}
\begin{figure}
\plotone{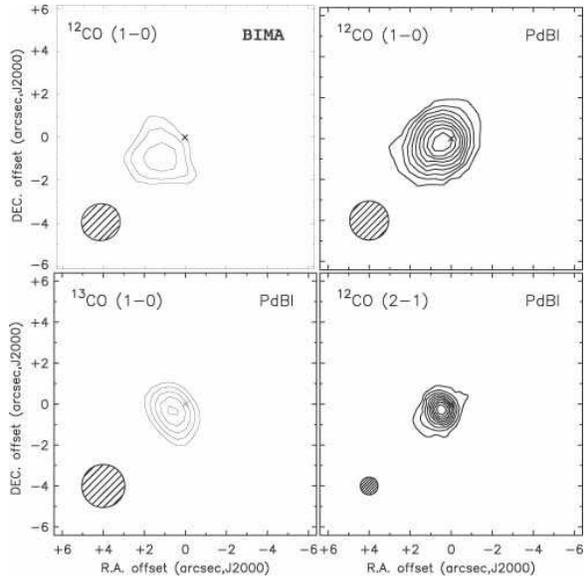}
\caption{{\it Top left:} BIMA image of the \twco(1-0) emission smoothed 
to a resolution of $2.0''$. The solid contours show the velocity
integrated emission between $-160$ and  240 km s$^{-1}$  in
$1\sigma$ steps (4~mJy~beam$^{-1}$) starting at $4\,\sigma$
(17\,mJy\,beam$^{-1}$). Dashed contours would show the corresponding
negative values for the flux density, if it were present in the
image. The ellipse in the lower left corner shows the synthesized
beam. To obtain integrated line fluxes, these and the following values should be multiplied by 
the velocity range. {\it Top right:} PdBI \twco(1-0) 0th moment image with $1.9''$
resolution, integrated between $-240$ and $+240$\,km\,s$^{-1}$. The
contours start at 10\% of the peak flux of 0.016 Jy beam$^{-1}$ and are in
steps of 10\%. {\it Bottom left:} PdBI \thco(1-0) image integrated over
the velocity range of $-240$ and $+240$\,km s$^{-1}$. The contours are at
0.3 ($3\sigma$), 0.4, 0.5, 0.7, and 0.8 mJy beam$^{-1}$ ($7\sigma$). {\it
Bottom right:} PdBI \twco(2-1) 0th moment image at $0.86''$ resolution,
integrated between $-240$ and $+240$ km\,s$^{-1}$. The contours start at
10\% of the peak flux of 0.017 Jy beam$^{-1}$ and are in steps of 10\%.
A cross marks the position of the optical center in each panel.}
\label{fig4}
\end{figure}
\begin{figure}
\plotone{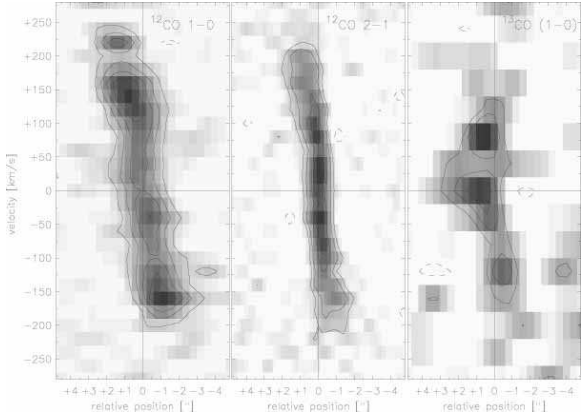}
\caption{PdBI data $pv$ diagrams along the major axis ($135\degr$) of I~Zw~1. The \twco(1-0) emission ({\it left}) is shown at $1.9''$
resolution, the \twco(2-1) emission ({\it middle}) at $0.86''$ resolution
and the \thco(1-0) line emission ({\it right}) at $2.1''$ resolution. The
contours start at $3\sigma$ in steps of $3\sigma$ for the \twco ~ data
with $1\sigma=2(1.7)$~mJy beam$^{-1}$ for the \twco(1-0) (\twco(2-1)) line.
For the \thco(1-0) the contours start at $2\sigma$ in steps of
$1\sigma$ with $1\sigma=0.3$~mJy beam$^{-1}$.} 
\label{fig5}
\end{figure}

\begin{deluxetable}{crrrr}
\tablecaption{I~Zw~1 \twco(1-0) beam averaged parameters \label{table1}}
\tablewidth{0pt}
\tablehead{
\colhead{Region} & \colhead{$\Delta v$}   & \colhead{$T_{B}$}   &
\colhead{$I_{\mathrm{CO}}$} &
\colhead{$N_{\mathrm{H}_2}$}  \\
\colhead{} & \colhead{[km s$^{-1}$]} &
\colhead{[K]}     & \colhead{[K km s$^{-1}$]}  &
\colhead{[$10^{21}$ cm$^{-2}$]}
} 
\startdata
Disk  ($9.0''$) \tablenotemark{a} &310 &0.03 &10  &2  \\
Nucleus ($2.0''$) &400 &1.8 &720 &144  \\
kpc ring ($0.7''$) &100 &12  &1200  &240
 \enddata
  \tablenotetext{a}{from \citet{schin98}}
 \end{deluxetable}

\subsection{ Spatial distribution of the \twco(1-0) line emission}

The gas disk of I~Zw~1 is clearly resolved in its \twco(1-0)
line emission at $0.7''$ resolution (Fig. \ref{fig2}).  These are the first molecular line
observations which resolve the emission in the nuclear region and show that
the emission is not centrally peaked: the strongest molecular line
emission is not coincident with the optical center of the galaxy, but
situated at a distance between $1''$ and $3''$, mainly to the South and
East of the galaxy center. This corresponds to a distance of between
1.2~kpc and 3.7~kpc from the nucleus. This emission, which occurs at
positive offsets in R.A. with respect to the major axis, has
predominantly positive velocities (see
position-velocity ($pv$) diagram in Figure \ref{fig3}).  When the velocity integrated
 BIMA \twco(1-0) data are smoothed to a resolution of $2''$ (Fig.~\ref{fig4}, top left),
which corresponds to the resolution of the corresponding  \twco(1-0) PdBI 
observations (Fig.~\ref{fig4}, top right), we find a positional offset of $\sim 1''$
in the center of emission of both images with respect to each other. The measured peak flux in the
BIMA image, which is smoothed to a resolution of $2"$
(Fig.~\ref{fig4}, top left), is 28~mJy beam$^{-1}$ or 11~Jy~beam$^{-1}\,\mathrm{km}^{-1}\,\mathrm{s}$, slightly larger than the
\twco(1-0) peak flux of 8 ~Jy~beam$^{-1}\,\mathrm{km}^{-1}\,\mathrm{s}$, at the same angular
resolution observed with the PdBI. The difference of $\sim 38\%$ is still within the expected range, if we assume an uncertainty of the PdBI calibration of $20$\%.

This significant spatial offset of the \twco(1-0) center of emission in the  BIMA and PdBI  observations  can not be due to a pointing error of the BIMA observations,
since a pointing error of this magnitude is not visible in the image
of the test quasar (Fig.~\ref{fig1}). The fact that the negative
velocity components in the BIMA observations (Fig.~\ref{fig3})
are less pronounced than
in the lower resolution PdBI data (Fig.~\ref{fig5}, left) indicates
that the molecular emission at these velocities has a more extended
distribution and therefore is partially resolved out in the high
resolution BIMA data. This effect naturally explains the shift
of the observed center in the BIMA data. Probably the most convincing
indication that the denser, and therefore probably more
spatially confined, molecular clouds are situated to the South-East of the
nucleus comes from the \thco(1-0) observations we obtained with the
PdBI (Fig.~\ref{fig4}, bottom left): the center of emission in these data
is also shifted towards the South-East. It is worthwile to note that the peak 
of the line emission in all CO 
transitions is shifted by $\sim 1''$to the South-East of the nominal 
optical center. This could be due to the fact that the distribution of the 
CO line intensity is asymmetric with respect to the AGN. This 
assumption could be tested if the position of the dynamical center could 
be accurately determined; however, even the resolution of our \twco(2-1) 
data is not adequate to do so.

\subsection{Properties of the molecular gas}

The observed $pv$ diagram along the major axis  at
$0.7''$ resolution (Fig.~\ref{fig3}) shows a circular rotation which is 
consistent with the lower 
resolution PdBI \twco(1-0) observations \citep[Fig.~\ref{fig5}, left; see also] [] {schar03}. The consistency of the observed $p-v$ diagrams in the BIMA and PdBI observations strongly supports the interpretation that the spatial offset of the peak emission in the BIMA  image is not an artifact due to calibration errors.  The peak emission of 35 mJy~beam$^{-1}$
in the BIMA $pv$ diagram, which is observed over a velocity interval
of about 100 km s$^{-1}$, corresponds to an average brightness
temperature of $T_B = 12$\,K in this velocity interval. This indicates
a large beam filling factor, respectively a significant amount
of molecular gas within a region of 600 pc$^2$. \citet{schin98}
 show that the standard conversion factor
$N(\mathrm{H}_2)/I(\mathrm{CO}) = 2\times 10^{20}\,
\mathrm{cm}^{-2} \,\mathrm{K}^{-1}\,\mathrm{km}^{-1}\,\mathrm{s}$ can be
applied to the observed \twco(1-0) line emission from I~Zw~1.

The peak flux of 18 mJy~beam$^{-1}$ in the velocity integrated $0.7''$ resolution image (Fig. \ref{fig2})  translates into a molecular hydrogen column density of $4.9\times 10^{23}$cm$^{-2}$. About 50\% of this emission has a velocity consistent with its origin being in a circumnuclear ring at a distance of $\sim~ 1.5''$. The corresponding column density of  $2.4\times 10^{23}$cm$^{-2}$  in the molecular ring is two orders of magnitude larger than the average molecular gas column density of $2\times10^{21} $cm$^{-2}$ in the disk region, and roughly a factor of two larger than the 
average column density of $1.4\times10^{23}$ cm$^{-2}$ in the nuclear region (the innermost $2''$ region: see Table~\ref{table1} for a summary of the observed \twco(1-0) molecular line properties). 

\subsection{Multi-transition analysis of the molecular line emission}

\begin{figure}
\plotone{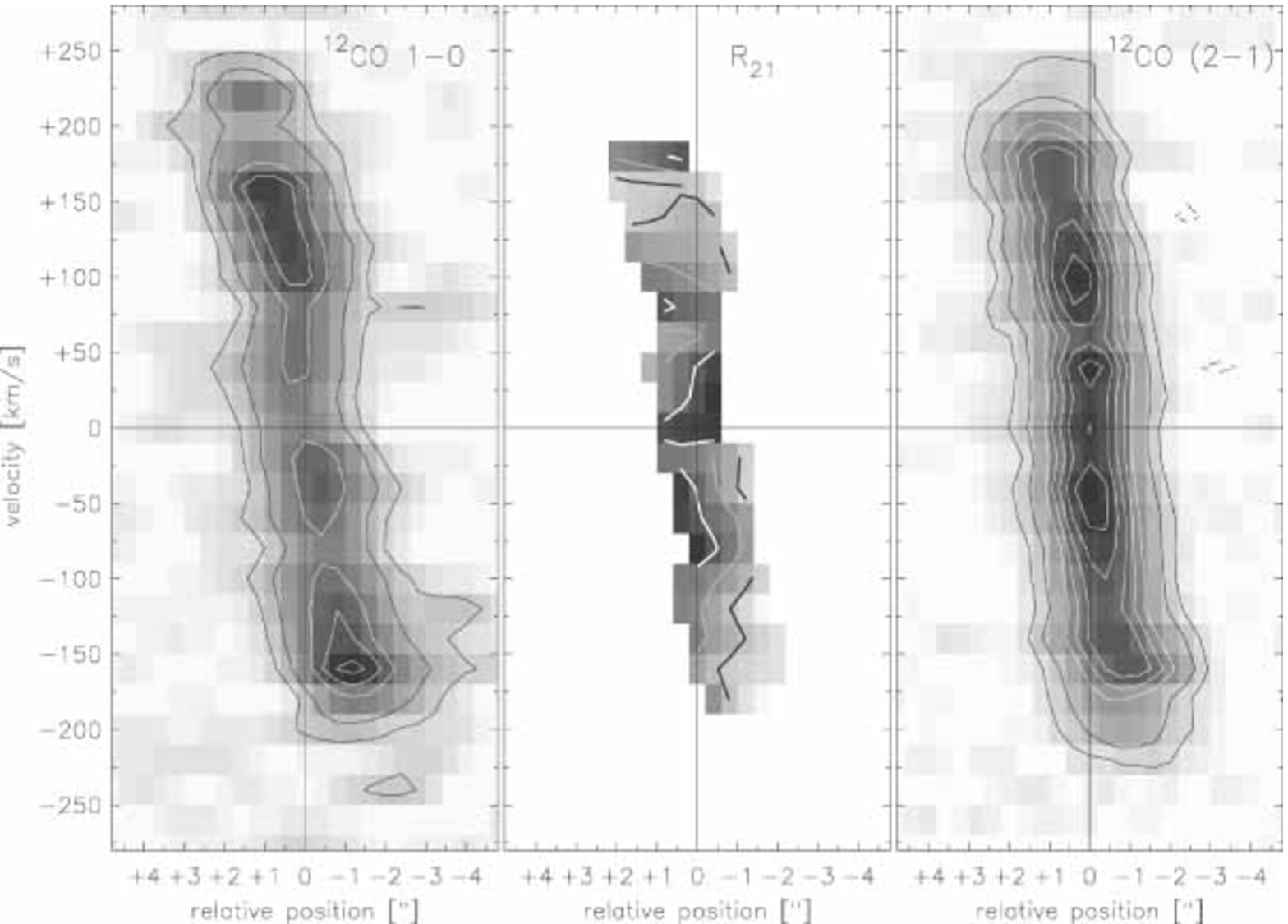}
\caption{PdBI \twco ~ line $pv$ diagrams at $2.1''$ resolution. The
$R_{21}=\frac{^{12}CO(2-1)}{^{12}CO(1-0)} $ peak brightness ratio ({\it
middle}) clearly shows a higher ratio in the inner $1.5''$ which is
decreasing by almost a factor of two towards higher velocities of $\pm
150\,$km s$^{-1}$. The contours are at 0.3 (black line), 0.4 (grey line) and
0.5 (white line). For reference the \twco(1-0) ({\it left}) and \twco(2-1) ({\it right}) $pv$
diagrams are shown at a resolution of $2.1''$. The contours start at
$3\sigma$ in steps of $3\sigma$ with $1\sigma$ of 1.8\,mJy beam$^{-1}$.
}
\label{fig:ratio}
\end{figure}

Our PdBI observations of the \twco(2-1) and \thco(1-0) line detect an
integrated line flux I$_{CO}$ of 18\,Jy\,km s$^{-1}$ and 0.5\,Jy\,km s$^{-1}$,
respectively. Thus they recover only about 30\% (15-20\%) of the observed
IRAM 30m single dish flux of 61\,Jy\,km s$^{-1}$ (3.1\,Jy\,km s$^{-1}$) for the
\twco(2-1) (\thco(1-0)) line \citep{eck94}. In contrast to the
BIMA \twco(1-0) observations, the $0.86''$ resolution PdBI \twco(2-1)
observations (Fig. \ref{fig4}, bottom right) do not show any
particular spatial structure in the nuclear region. This discrepancy can only be explained by a difference in the excitation of the molecular gas in the nucleus (innermost $1''$) and the emission from molecular clouds which are situated further out. This interpretation is supported by both the PdBI  and BIMA \twco(1-0)  observations:
A comparison of the $pv$ diagrams from the PdBI \twco(1-0)
observations from \citet{schin98} (here chosen for the better
signal to noise) and our new PdBI \twco(2-1) and \thco(1-0)
observations can be seen in Figure~\ref{fig5}. The peaks of the line
intensity are shifted from $\pm 150\,$km\,s$^{-1}$ in the \twco(1-0) line to
lower velocities of $\pm (50-100)\,$km\,s$^{-1}$ in the \twco(2-1) line. This
difference becomes even more obvious in the
$R_{21}\,=\,\frac{^{12}CO(2-1)}{^{12}CO(1-0)}$ line peak ratio map of the
major kinematic axis $pv$ diagram shown in Figure \ref{fig:ratio}. For these ratios the resolution of the \twco(2-1) observations was smoothed to match the PdBI \twco(1-0) observations. The
line peak brightness ratio in the inner $\sim 1.5''$ of I~Zw~1 is
$\approx 0.5$. This drops by almost a factor of two to $\approx 0.3$
at the position of the
\twco(1-0) line peaks at $\pm 150\,$km$\,s^{-1}$. This suggests that the outer
\twco(1-0) clouds, which are apparently not resolved out  by the BIMA observations with a similar angular resolution as the PdBI \twco(2-1) observations, are more 
subthermally excited than the molecular gas in the innermost region of I~Zw~1. 
The fact that the denser and more thermalized molecular
material breaks up into several distinct velocity components in the \twco(2-1)
emission  (Figure~\ref{fig5}) strongly suggests that
circumnuclear star formation and not the central AGN is exciting the
molecular line emission.

The derived $r_{10}=\frac{^{12}CO(1-0)}{^{13}CO(1-0)}$ line peak brightness ratio for
the three \thco(1-0) line peaks above $3\sigma$ (see Figure \ref{fig5},
right) is about 15. This value can probably be regarded only as an upper limit due to the large amount of missing flux which has been resolved out by our interferometric observations,  in particular in the \thco(1-0) observations (see above). We therefore avoid further interpretation of these ratios.

NIR spectra of the inner $3.3''$ of I~Zw~1 indicate the presence of
circumnuclear star formation (Schinnerer et al. 1998). The millimeter line ratios
for the inner $1.5''$ suggest that these GMCs could be the site of this
massive star formation. However, the  BIMA observed \twco(1-0) GMCs,
which are  further out, are also likely candidates for the observed
starburst due to their high column densities. A starburst in these clouds might
be triggered by the interaction of I~Zw~1 with the companion galaxy \citep[see] [] {schar03}.

\subsection{Molecular gas kinematics}

\begin{figure}
\plotone{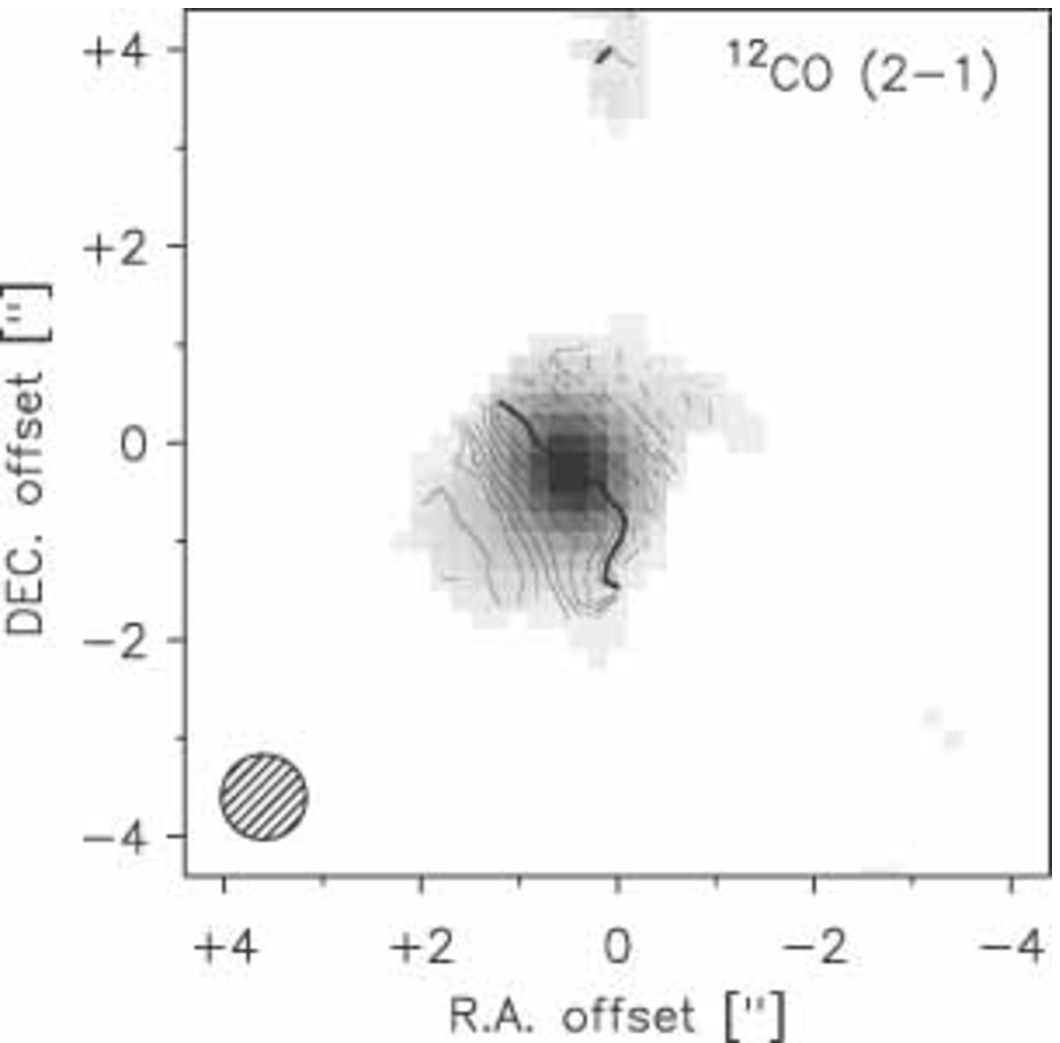}
\caption{PdBI \twco(2-1) velocity field at 0.86'' resolution.
The velocity field is shown in contours (the thick line represents
$v_{hel}=18,316\,$km s$^{-1}$, the solid (dashed) lines are positive
(negative) velocities in steps of 20\,km s$^{-1}$). The 0th moment image is
shown in grayscale for comparison. The beam is indicated in the lower
left corner.
}
\label{fig:co21vel}
\end{figure}

There is some enhanced emission in the BIMA $pv$ diagram
(Fig.~\ref{fig3}) at velocities which are ''forbidden'' with respect to
circular motion, most prominently at a positive angular offset of $2''$ along
the major axis between ${-40}$ and ${-60}$ km\,s$^{-1}$. Other than the
emission from the circular rotation, it is pronounced only in
one bin of the {\it p-v} diagram, and with a signal to noise ratio of
$4\sigma$ its significance is not sufficient to be considered a clear
detection. The \twco(2-1) velocity field (Fig. \ref{fig:co21vel}) also
shows indications of deviation from pure circular rotation for radii
of $r \ge 1''$, as is evident in the change in the line of nodes. Due
to the lack of emission, this potential non-circular motion can not be
followed to larger radii. Together with the ``forbidden'' \twco(1-0)
it can be seen as circumstantial evidence that some transportation
mechanism for the molecular gas is active in the disk of
I~Zw~1. However, data of higher angular resolution and improved 
sensitivity will be
needed to confirm this. A more detailed analysis of the observed
rotational pattern can be found in \citet{schar03}.

\section{Summary and Conclusions}

Our BIMA \twco(1-0) observations represent the first molecular line
observations of the host galaxy of a QSO with sub-kpc resolution. The
molecular gas disk of a QSO is for the first time spatially resolved
into individual giant molecular cloud complexes. We observe giant
molecular clouds with peak molecular hydrogen column densities of
$2.4\times 10^{23}$cm$^{-2}$ at a distance between $1''$ and $3''$ from
the nucleus. With such high column densities these clouds could be
actively forming stars.

The combination of BIMA observations with new and previously published 
PdBI CO data allows for the first interferometric multi-transition study in the
disk of I~Zw~1. The line brightness ratios in the inner $1.5''$ are
consistent with moderately dense cold GMCs, and they are not peaked at
the center. This strongly suggests that the AGN has no significant
effect on the central molecular material.

The exact location of the circumnuclear starburst very likely seen in
nuclear NIR spectra (Schinnerer et al. 1998) is still not
identified. Possible candidates are the resolved \twco(1-0) molecular
clouds seen by BIMA as well as the more thermalized material traced by
the higher line ratios inside the inner $1.5''$. However, any of these
molecular clouds are likely sites of a massive starburst which
contributes to the observed far-infrared luminosity of
$L_{FIR}=10^{11.3}L\sun$. 

The distribution and properties of the molecular gas in the host
galaxy of the nearby QSO I~Zw~1 are quite similar to what is observed
in nearby low luminosity AGNs \citep[e.g.] [] {pag01, gar03}. 
This suggests that the ISM in QSO at high redshift might be
similar to nearby low luminosity AGN as well.

Four additional attempts to re-observe I~Zw~1 in \twco(1-0) with BIMA
in the A configuration all failed to provide useful data, due to
insufficient phase coherence of the observations. This demonstrates
that the observations presented here are on the limit of what can be
done without active phase correction systems, at least from a site
such as Hat Creek where the BIMA array is situated.

\acknowledgments

We would like to thank P.J. Teuben for his helpful advice with the data reduction. The constructive comments from the anonymous referee are much appreciated.
A. Eckart. and J. Scharw\"achter are supported by DFG grant SFB 494.  J. Scharw\"achter is also supported by a scholarship for doctoral students of the ÒStudienstiftung
des deutschen VolkesÓ.

\end{document}